# The Structure of Dark Matter Haloes in Dwarf Galaxies

A. Burkert

Max-Planck-Institut für Astronomie

Königstuhl 17, D-69117 Heidelberg, Germany

astro-ph/9504041    12 Apr 95

# ABSTRACT

Recent observations indicate that dark matter haloes have flat central density profiles. Cosmological simulations with non-baryonic dark matter predict however self-similar haloes with central density cusps. This contradiction has lead to the conclusion that dark matter must be baryonic. Here it is shown that the dark matter haloes of dwarf spiral galaxies represent a one-parameter family with self-similar density profiles. The observed global halo parameters are coupled with each other through simple scaling relations which can be explained by the standard cold dark matter model if one assumes that all the haloes formed from density fluctuations with the same primordial amplitude. We find that the finite central halo densities correlate with the other global parameters. This result rules out scenarios where the flat halo cores formed subsequently through violent dynamical processes in the baryonic component. These cores instead provide important information on the origin and nature of dark matter in dwarf galaxies.

*Subject headings:* Dark matter — galaxies: spiral — galaxies: structure

## 1. Introduction

The nature of the dark matter (DM) is one of the most important and still unsolved problems in astronomy. Current cosmological models assume that there exists a nonbaryonic, cold dark matter (CDM) component which consists of non-relativistic particles that are assumed to be collisionless and to interact with the baryons only through gravitational forces. CDM simulations indeed reproduce quite well the observed clustering on galactic scales (Davis et al. 1985). They also predict the existence of extended dark matter haloes around galaxies (Frenk et al. 1985) with constant outer rotation curves, in agreement with the observations (Casertano & van Gorkom 1991).

In order for the galactic gas and stars to achieve a constant circular velocity in the outer regions, the density profile of the DM halo must fall approximately proportional to $r^{-2}$ in the relevant radius range. Such a profile resembles the density structure of an isothermal, self-gravitating system of particles which is characterized by a constant velocity dispersion. DM haloes are therefore often approximated by a so called modified, isothermal density profile (Begeman et al. 1991)

$$\rho(r) = \frac{\rho_0}{1 + \frac{r^2}{r_c^2}} \quad (1)$$

where $r_c$ is the core radius and $\rho_0$ is the central dark matter density.

Early cosmological N-body calculations did not have enough resolution to resolve the inner structure of DM haloes. Recent CDM-simulations with much higher spatial resolution (Dubinski & Carlberg 1991, Navarro et al. 1995) however lead to DM haloes with an inner density distribution that is inconsistent with equation 1. Whereas the isothermal density profile becomes almost constant at radii smaller than $r_c$ and has a finite central density, the numerical models indicate a density distribution which diverges as $\rho \sim r^{-1}$ in the inner parts, leading to an infinite density in the center. Unfortunately such central density cusps are hard to verify in normal spiral galaxies as their inner parts are gravitationally dominated by baryons. The inferred DM properties then depend strongly on the assumed mass-to-light ratios of the galactic disk and of the spheroidal central stellar component. The situation is different in dwarf spiral systems which recently have been observed with high resolution. It has been shown that some of these systems are completely dominated by dark matter (Carignan & Freeman 1988). They therefore represent ideal candidates for an investigation of the inner structure of dark matter halos. Flores & Primack (1994) and Moore (1994) recently showed that the DDO galaxies have rotation curves which rule out singular halo profiles with infinite central densities and are in good agreement with the modified isothermal density law (eq. 1). The origin of these shallow DM cores represents an interesting and challenging cosmological puzzle which is not yet solved.

## 2. The Universal Density Profiles of Dark Matter Haloes

Scale free cosmological scenarios with non-baryonic, dissipationless DM predict that all DM haloes should have similar structure (Navarro et al. 1995). This prediction is tested in Figure 1 which shows the DM mass distribution as derived from the observed neutral hydrogen rotation curves of four well studied dwarf spiral galaxies. These systems are completely DM dominated, even in their innermost regions. Their rotation curves therefore trace directly the DM mass distribution and the uncertain gravitational contribution of the baryonic component can be neglected. All four profiles indeed follow the same universal mass relation, implying a universal halo structure. Note that the modified, isothermal profile (dashed curve) only fits the inner regions; it rises somewhat too fast at large radii. The dotted and dot-dashed curves show the mass distribution as predicted from cosmological CDM simulations on cluster scales (Navarro et al. 1985). These calculations predict DM profiles which are given by the fitting function $\rho(r) = 300\,\bar{\rho}\,r_{200}^3 \times r^{-1}(\,r + 0.25\,r_{200}\,)^{-2}$ where $\bar{\rho}$ is the mean density of the universe at the epoch when the DM halo formed. $r_{200}$ denotes the radius of the DM sphere of mean overdensity of 200. These profiles predict too much mass at small radii as a result of the central density cusp (Flores & Primack 1994, Moore 1994).

The observed universal mass profiles can be fitted nicely over the whole observed radius range by the following, purely phenomenological density distribution:

$$\rho_{DM}(r) = \frac{\rho_0 r_0^3}{(r+r_0)(r^2+r_0^2)}. \qquad (2)$$

which is shown in Figure 1 by the solid line. $\rho_0$ and $r_0$ are free parameters which represent the central density and a scale radius, respectively. This revised density law resembles an isothermal profile in the inner regions ($r \ll r_0$) and predicts a finite central density $\rho_0$. $r_0$ is similar to the core radius $r_c$ of the isothermal fit formula (equation 1). For large radii, the mass distribution of the isothermal profile would diverge proportional to r. Equation 2 leads to mass profiles which diverge logarithmically with increasing r, in agreement with the predictions of cosmological CDM calculations (Navarro et al. 1995).

## 3. The Dark Matter Scaling Relations

Supposing that the DM haloes of dwarf spirals are indeed all similar with a density structure as described by equation 2, we next can investigate the question whether there exists a relation between $\rho_0$ and $r_0$. Instead of studying $\rho_0$, which cannot be observed directly, we choose as free parameter the rotational velocity $v_0$ of the DM rotation curve at $r_0$. $r_0$ and $v_0$ are determined by fitting the theoretically predicted rotation curve (using equation 2) to the observed DM rotation curves. Accurate neutral hydrogen rotation curves are now available for about a dozen low-mass disk galaxies. From this sample we select those candidates which are clearly dark matter dominated at $r_0$. This restriction considerably reduces the error in determining $v_0$ due to uncertainties in the mass-to-light ratios of the gaseous and stellar components. In addition, the inner DM profiles are not strongly affected by the gravitational force of the baryons. Only those galaxies can be used for which the rotation curve is known at $r_0$. These constraints reduce the sample to seven galaxies. The present sample consists of the four previously studied galaxies plus three systems (Puche & Carignan 1991, Broeils 1992): NGC55, NGC300, and NGC1560. For these galaxies the gravitational force of the visible components is not completely negligible inside $r_0$. The contribution of the visible matter to the gravitational potential therefore had to be subtracted, leading to a residual DM rotation curve. Here we adopt the best three component fits to the rotation curves as suggested by Puche & Carignan (1991) for NGC55 and NGC300 and by Broeils (1992) for NGC1560.

Figure 2 shows $v_0$ of all galaxies as a function of $r_0$. We find a strong, linear correlation between $r_0$ and $v_0^{1.5}$. Assuming spherical symmetry, one can derive the following scaling relations for $v_0$, the total dark matter mass $M_0$ inside $r_0$, the central density $\rho_0$ and $r_0$:

$$\begin{aligned} v_0 &= 17.7 \left(\frac{r_0}{kpc}\right)^{2/3} \frac{km}{s} \\ M_0 &= 7.2 \times 10^7 \left(\frac{r_0}{kpc}\right)^{7/3} M_\odot \\ \rho_0 &= 4.5 \times 10^{-2} \left(\frac{r_0}{kpc}\right)^{-2/3} \frac{M_\odot}{pc^3} \end{aligned} \quad (3)$$

The equations 3 demonstrate, that the density profiles of the dark matter haloes in dwarf galaxies depend only on one free parameter, e.g. $r_0$. Given $r_0$, the central density $\rho_0$ (equation 3) and by this the complete DM density profile (equation 2) is specified. One should note, that Moore (1994) also noted a trend of increasing size of the galaxy with the peak rotational velocity $v_{max}$. Unfortunately, $v_{max}$ is not well determined in most systems as the rotation curves tend to rise past the observed region. Moore's conclusion is however in qualitative agreement with the present results.

## 4. Discussion of the Scaling Relations

From the scaling laws (equation 3) one can derive the following relation: $M_0 \sim v_0^{3.5}$. This relation is similar to the Tully-Fisher-Relation (TFR) for massive spiral galaxies (Tully & Fisher 1977) which predicts that the galaxies' luminosity L correlates with their maximum rotational velocity $v_{max}$ according to $L \sim v_{max}^4$. Note however that the TFR combines the mass of the luminous galactic components with the maximum of the rotation curve whereas here, the velocity of the DM rotation curve at $r_0$ is correlated with the DM mass inside $r_0$. There exists no obvious reason why the relations should be similar. In addition, it has been shown that the TFR is not valid for DM-dominated DDO galaxies (Carignan & Freeman 1988) as in these systems the baryonic component is not massive enough in order to determine $v_{max}$.

Like the TFR, the equations 3 can be used as a tool for determining the distances to DM-dominated dwarf galaxies, provided their HI-rotation curves are known up to $r_0$. In this case, $r_0$ in arcsec can be specified by fitting the universal velocity profile as given by equation 2 to the observed rotation curve. $v_0$ can then be determined directly from the data. Equation 3 then gives the corresponding scale radius $r_0$ in units of kiloparsecs. A comparison with the observed radius in arcsec leads to the distance of the galaxy.

There might exist a cosmological explanation for the scaling relations (equation 3). Consider a spherical region with DM density $\rho$ and overdensity $\overline{\delta} = \rho/\overline{\rho} - 1$ in an otherwise uniform Einstein-de Sitter universe with mean density $\overline{\rho}$. The mass of this DM halo is $M_h = 4\pi\rho R^3/3$ where R is its physical radius. The density fluctuation first expands with the universe, reaches a maximum radius $R_m$ at a certain redshift $z_m$ and then collapses again. For a collisionless system of DM particles, this DM halo will achieve a virial equilibrium state with a final radius $R_h = 0.5 R_m$. Bardeen et al. (1986) and White (1994) demonstrate in detail how one can calculate $R_m$ and $z_m$ as a function of $M_h$ for certain cosmological scenarios. As an example, the solid line in Fig. 3a shows the theoretically predicted mass-radius relation for virialized, dissipationless CDM haloes which formed from primordial $1\sigma$ density fluctuations in the standard CDM-universe.

Let us now assume that all the observed DM haloes result from $1\sigma$ fluctuations. As all haloes are self-similar there should exist a universal, linear relation between the observed scale parameters $(M_0, r_0)$ and the virial parameters $(M_h, R_h)$. In order to determine this relation, the dotted curve in Fig. 3a shows the observed mass profile M(r) for the DM halo of the dwarf galaxy DDO154. Its scale parameters $(M_0, r_0)$ are shown by the filled circle. If DDO154 formed from a $1\sigma$ fluctuation, its virial parameters should lie somewhere on the solid line. The virial parameters are therefore determined by the intersection (open circle) of the dashed and solid curve, leading to the following relations

$$R_h = 3.4 \times r_0 \qquad (4)$$
$$M_h = 5.8 \times M_0$$

Supposing that the DM haloes of all the observed dwarf spirals are similar and result from $1\sigma$ CDM fluctuations, their cosmological virial parameters $M_h$ and $R_h$ can be derived from the observed scale parameters $r_0$ and $M_0$, using the equations 4. Figure 3b shows $M_h$ versus $R_h$ for all the dwarf spirals of our sample. The data points indeed follow very nicely the theoretical mass-radius relation.

## 5. Conclusions

As demonstrated in the previous section, cosmological models with non-baryonic, dissipationless cold dark matter can in principle explain the scaling relations of DM haloes around dwarf galaxies. The dotted and dashed lines in Figure 3b show however clearly that a much larger spread of the virial parameters is expected as regions with different primordial overdensities could in principle form low-mass DM haloes. If the proposed cosmological interpretation is valid we have to draw the conclusion that only fluctuations with a certain, fixed amplitude managed to form dwarf spiral galaxies with dominating DM haloes. This interesting question will be addressed in greater details in a subsequent paper.

Cosmological models predict self-similar haloes, in agreement with the present observations. Figure 1 shows however that numerical CDM simulations lead to DM density profiles with shapes that disagree with the observations. A possible answer to this problem might be dynamical processes within the baryonic, visible component (Flores & Primack 1994, Moore 1994) which affect the gravitational potential in the inner regions of DM haloes, by this leading to a secular change of the DM density profile. Ejection of gas might for example decrease the gravitational force in the inner region, triggering an expansion of the inner DM halo and leading to a finite central density with an almost constant density core. Although such a scenario seems at first very attractive, the present results indicate that significant fine tuning would be required between the early cosmological, collisionless collapse phase which leads to DM haloes in the first place and the secular energetic processes in the baryonic matter which occur later. It is unlikely that DM haloes have self-similar density profiles, if their inner structure is subsequently changed by these completely different, secular processes, whereas the outer regions result from the collisionless relaxation phase of the DM halo. We rather would expect to find a second independent parameter which describes the structure of the cores of DM haloes and is determined by

10the efficiency with which the baryonic processes affect the inner DM halo.

The observed shallow central density profiles of DM haloes are more likely a direct result of their collisionless formation history. If DM, for example, has a finite maximum phase space density $f_{max}$ (as in the case of warm or hot dark matter) conservation of $f_{max}$ during the collisionless halo formation phase limits the halo's core phase space density $f_0$, leading to a central region with $f_0 \approx f_{max}$. If the central velocity dispersion $\sigma_0$ of the DM halo is finite, this constraint implies a finite maximum central density $\rho_0 \approx f_{max} \times \sigma_0^3$. This scenario would however require that the central phase space densities of DM haloes are similar which can be ruled out from the scaling relations. According to the equations 3, $f_0$ scales as $f_0 \sim \rho_0/v_0^3 \sim r_0^{-8/3}$ and therefore decreases steeply with increasing scale length. A finite maximum phase space density therefore is unlikely to be the correct explanation.

The nature of DM and the origin of DM haloes with finite central densities remains ill understood. The discussed discrepancies between CDM calculations and the observations indicate that some important physical features, which are related to the nature and origin of dark matter are still missing in cosmological models.

I would like to thank Simon White for interesting and critical discussions and the referee for clarifying comments.

---

FIGURE CAPTIONS

FIG. 1: Dark matter mass profiles, derived from the observed rotation curves are shown for the following dwarf spiral galaxies: DDO154 (open triangle, Carignan & Beaulieu 1989), DDO105 (open square, Schramm 1992), NGC3109 (open circle, Broeils 1990) and DDO170 (starred, Lake et al. 1990). The scale radius $r_0$ is defined by equation 2. The errorbar at the innermost triangle represents the observational uncertainty in determining the rotational velocity and the mass at these radii. For larger radii, the errorbars are as large as the size of the symbols. The isothermal fit is shown as dashed curve, assuming a core radius $r_c = r_0$. The solid line shows the revised profile (equation 2). The dotted and dot-dashed curves show DM profiles as predicted from CDM calculations. The best fit to the data (dotted line) is achieved if one assumes $r_{200} = 17.5 \times r_0$ which for an $\Omega = 1, h = 0.5$ CDM universe corresponds to a formation redshift of $z = 0.6$. The dot-dashed curve shows the mass profile as predicted from the CDM calculations if $r_{200} = 5 \times r_c$ with a corresponding formation redshift of $z = 1.5$.

FIG. 2: The scaling relation between the rotational velocity $v_0$ of the DM haloes at $r_0$ is shown as function of the scale radius $r_0$ (equation 2). The open circles show the four best known DDO galaxies which have been used in Figure 1. The filled circles show three additional galaxies for which the contribution of the baryonic component to the rotation curve at $r_0$ had to be subtracted. The error bars show the observational uncertainty in determining $v_0$ as quoted in the literature. The dashed line shows a fit (equation 3) through the data points. An extension of the data to larger radii fits nicely the halo parameters of Sc-Im galaxies (Kormendy & Sanders 1992). For comparison, the dotted curve shows a linear fit $v_0 \sim r_0$.

FIG. 3: The solid line in Figure 3a shows the predicted mass-radius relation $M_h(R_h)$ for virialized, cold dark matter haloes which formed from primordial $1\sigma$ density fluctuations in the standard ($\Omega = 1$; h = 0.5)-universe with a biasing factor of $b = 0.6$ (White et al. 1993). The dotted line shows the halo mass profile for DDO154 as derived from its rotation curve and extended to larger radii, using equation 2. The filled circle indicates the location of the observed halo scale parameters $(M_0, r_0)$, the open circle shows the predicted virial parameters $(M_h, R_h)$ for DDO154.

Figure 3b compares the virial parameters $M_h$ versus $R_h$ of the observed DM haloes (open circles) with the mass-radius relation as predicted for a standard CDM universe. The dotted, solid and dashed lines represent primordial $0.5\sigma$, $1\sigma$ and $2\sigma$ perturbations, respectively.

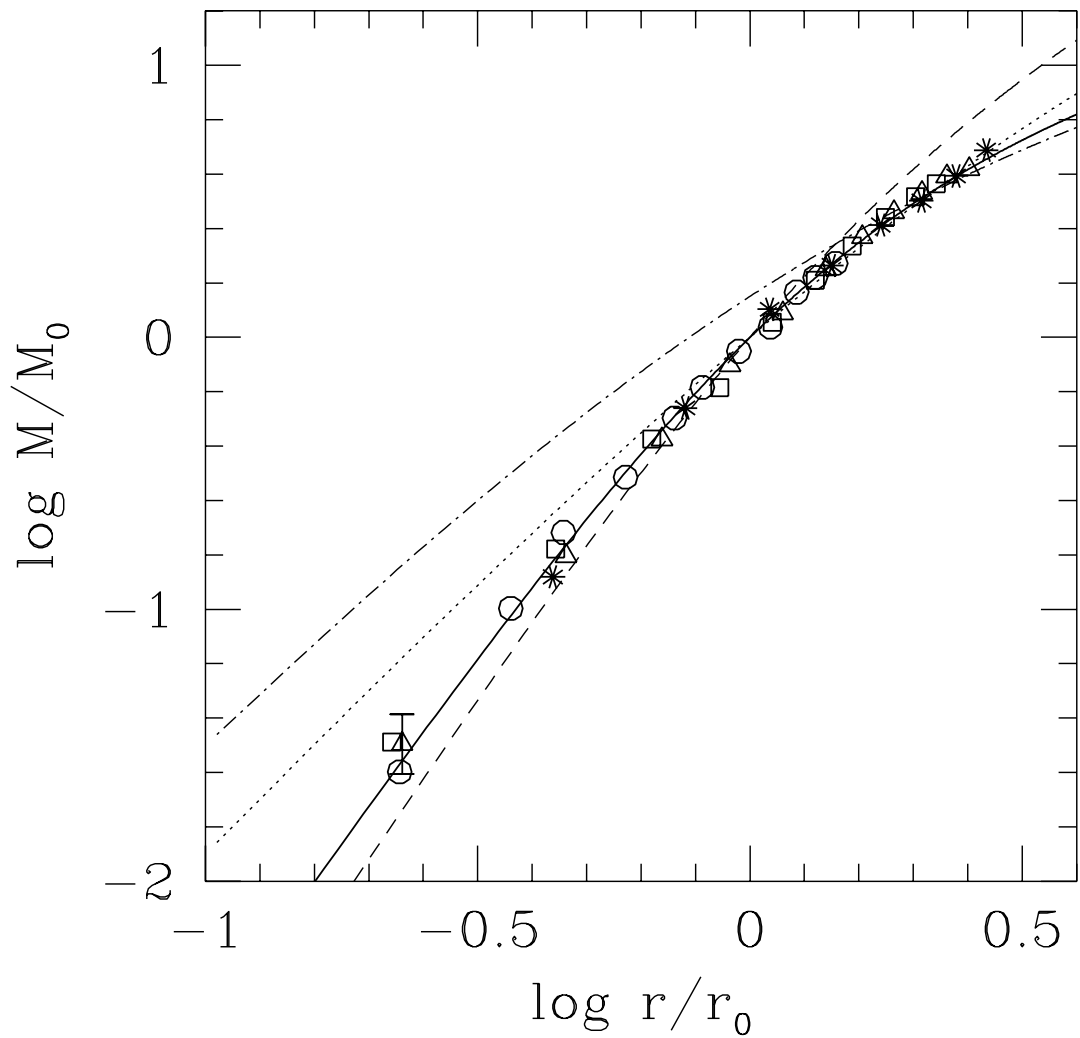

Fig. 1.—

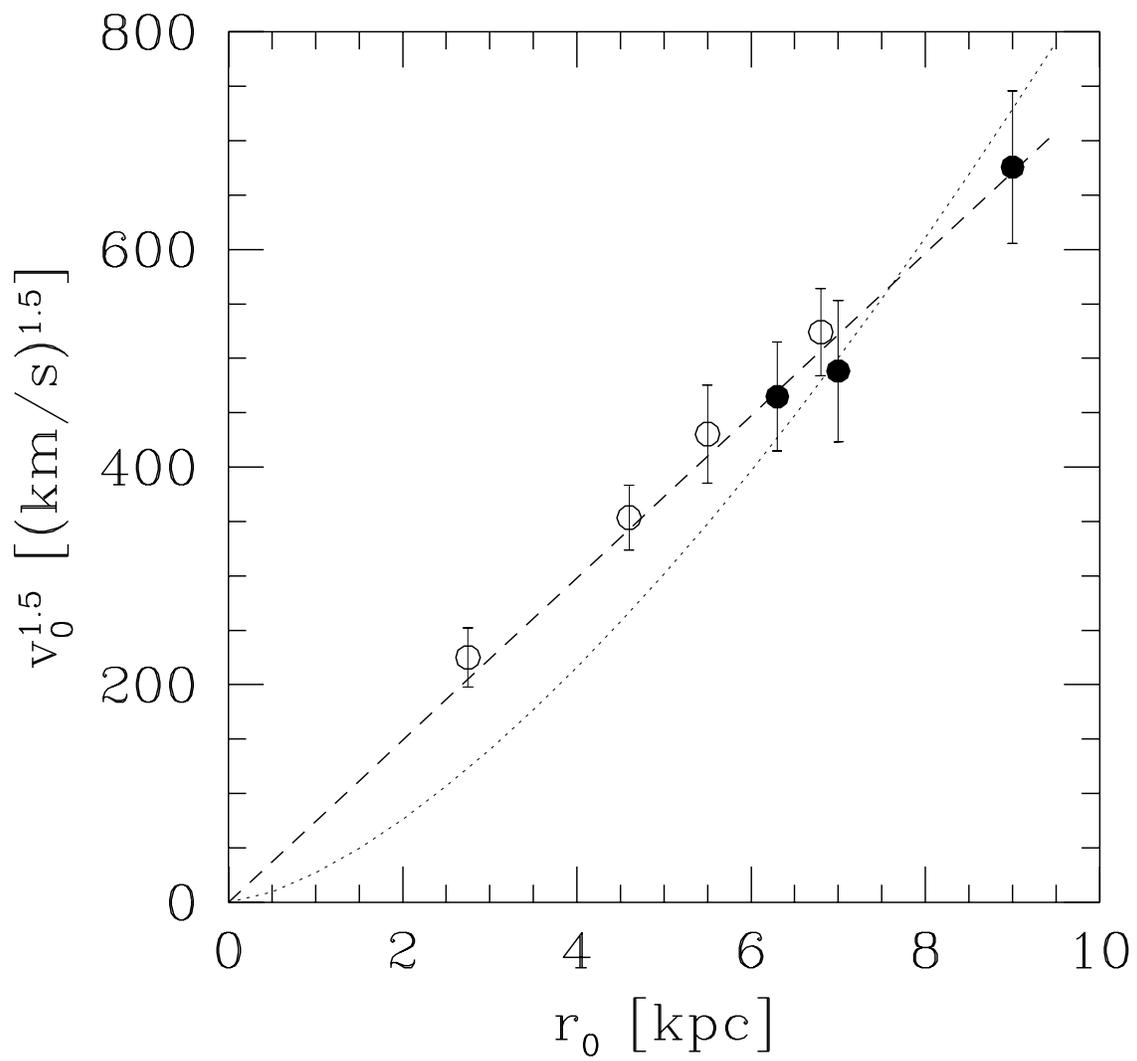

Fig. 2.—

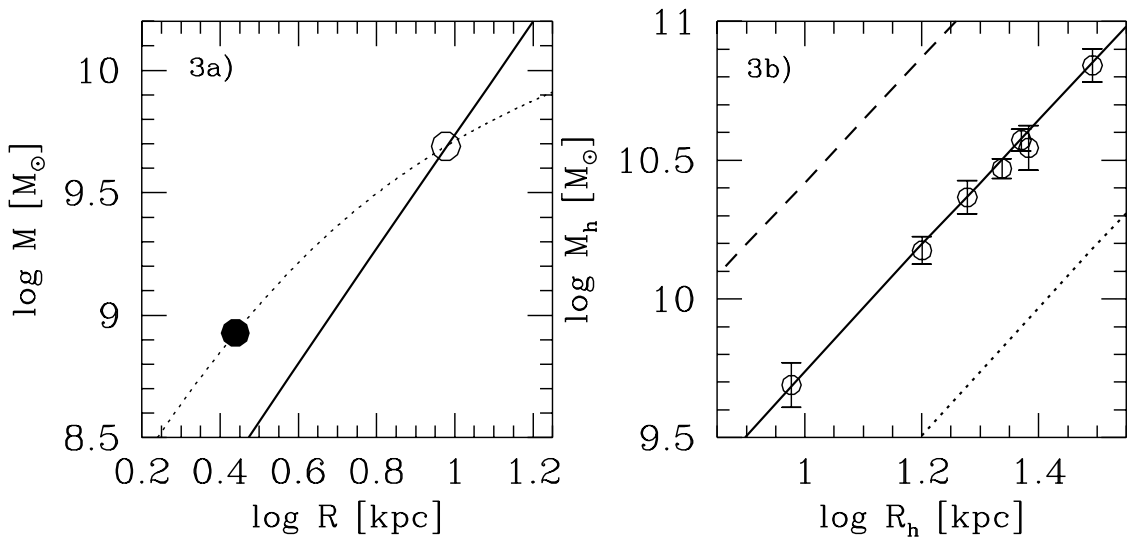

Fig. 3.—